\documentclass[12pt]{article}
\usepackage{amsmath}
\usepackage{amssymb}
\usepackage{citesort}
\usepackage{graphicx}
\usepackage{pstricks}
\newcommand{\BE}{\begin{equation}}
\newcommand{\EE}{\end{equation}}
\def\bea{\begin{eqnarray}}
\def\eea{\end{eqnarray}}
\begin{document}

\setcounter{page}{0} \thispagestyle{empty}


\vspace*{1.5cm}

\begin{center}
{\large \bf Is the standard singlet Higgs a true massive field ?}

\end{center}

\vspace*{1.5cm}

\renewcommand{\thefootnote}{\fnsymbol{footnote}}

\begin{center}
{\large M. Consoli}
 \\[0.3cm]
 {INFN - Sezione di Catania, I-95123 Catania, Italy}
\end{center}

\vspace*{0.5cm}

{

\vspace*{1.0cm}

\renewcommand{\abstractname}{\normalsize Abstract}
\begin{abstract}
The phenomenon of spontaneous symmetry breaking admits a physical
interpretation in terms of the Bose-condensation process of
elementary spinless quanta. In a cutoff theory, this leads to a
picture of the vacuum as a condensed medium whose excitations might
deviate from exact Lorentz covariance in both the ultraviolet and
infrared regions. For this reason, the conventional singlet Higgs
boson, the shifted field of spontaneous symmetry breaking, rather
than being a purely massive field, might possess a gapless branch
describing the long-wavelength fluctuations of the scalar
condensate. To test this idea, that might have substantial
phenomenological implications, I compare with a detailed lattice
simulation of the broken phase in the 4D Ising limit of the theory.
The results are the following: i) differently from the symmetric
phase, the single-particle energy spectrum is not reproduced by the
standard massive form ii) for the value of the hopping parameter
$\kappa=0.076$, increasing the lattice size from $20^4$ to $32^4$,
the mass gap  is found to decrease from the value 0.392(1) reported
by Balog et al. (see Nucl. Phys. {\bf B714} (2005) 256) to the value
0.366(5). Both results confirm that, in the infrared region, the
standard singlet Higgs cannot be considered as a simple massive
field. Several arguments indicate that, approaching the continuum
limit of the lattice theory, the observed volume dependence of the
mass gap might require larger and larger lattice sizes before to
show up.

\end{abstract}

\vspace*{0.5cm}
\begin{flushleft}
\end{flushleft}
\renewcommand{\thesection}{\normalsize{\arabic{section}.}}
\vfill\eject
\section{\normalsize{Introduction}}
\label{Introduction}

The idea of a `condensed vacuum' is generally accepted in modern
elementary particle physics. Indeed, in many different contexts, one
introduces a set of elementary quanta whose perturbative empty
vacuum state $|o\rangle$ is not the true ground state of the theory.
In the physically relevant case of the Standard Model, where the
condensation process corresponds to spontaneous symmetry breaking
(SSB) through the $\lambda\Phi^4$ sector, the situation can be
summarized saying \cite{thooft} that "What we experience as empty
space is nothing but the configuration of the Higgs field that has
the lowest possible energy. If we move from field jargon to particle
jargon, this means that empty space is actually filled with Higgs
particles. They have Bose condensed". The translation from field
jargon to particle jargon can be obtained, for instance, along the
lines of Ref.\cite{mech} where the substantial equivalence between
the one-loop effective potential of quantum field theory and the
energy density of a dilute particle gas was established.

With these premises, it was argued in Ref.\cite{consoli} that such a
physical representation of the vacuum might run into a contradiction
with the conventional picture where the standard singlet Higgs
boson, the shifted field of SSB, is represented as a purely massive
field. In fact, in the hydrodynamical regime of disturbances whose
wavelengths are much larger than the mean free path of the
elementary constituents, quite independently of the Goldstone
phenomenon and even for a spontaneously broken one-component
$\lambda\Phi^4$ theory, one would expect that the lowest excitations
of the system correspond to small displacements of the condensed
quanta that already preexist in the ground state. In this case, for
momenta ${\bf p}\to 0$, the energy spectrum should end with an
acoustic "phonon" branch $\sim c_s|{\bf p}|$ as it happens in
superfluids and in all known condensed media. The existence of such
a regime is, in fact, a very general result \cite{stevenson2001}
that does not depend on the details of the short-range interaction
and even on the nature of the elementary constituents. For instance,
the same coarse-grained description is found in superfluid fermionic
vacua \cite{voloreport} that, as compared to the Higgs vacuum, bear
the same relation as superfluid $^3$He has to superfluid $^4$He.

In this sense, as discussed in Ref.\cite{consoli2}, the idea of a
massive excitation spectrum of the form $\sqrt { {\bf p}^2 + M^2_h}$
down to ${\bf p}=0$ can be considered equivalent to the
incompressibility limit where there is no phonon branch and
$c_s=+\infty$. This can be a valid representation of the vacuum for
the continuum theory where, as in axiomatic quantum field theory,
one can derive the exact Lorentz covariance of the energy spectrum.
However, in the condensed phase of a cutoff theory one is faced with
"reentrant violations of special relativity in the low-energy
corner" \cite{volo} that only disappear in the continuum limit where
the ultraviolet cutoff $\Lambda \to \infty$.

In a condensate of spinless particles, these violations would be
simple density fluctuations extending over a small region $|{\bf p}|
< \delta$. Lorentz covariance can become exact if the shell $\delta$
vanishes, in units of the scale $M_h$ associated with the massive
part of the spectrum, in the limit $\Lambda \to \infty$. Introducing
dimensionless units, $\epsilon\equiv {{\delta}\over{M_h}}$ and
$t\equiv {{\Lambda}\over{M_h}}$, the "reentrant" nature of the
Lorentz violating infrared effects implies the equivalence of the
$\epsilon \to 0$ and $t\to \infty$ limits as, for instance, embodied
in the single relation \cite{consoli2} \BE \label{golden} \delta\sim
{{M^2_h}\over{\Lambda}} \EE

In the cutoff theory, the very long-wavelength phonons of the Bose
condensate would produce a very weak $1/r$ potential \cite{grifols}.
Thus one can consider \cite{consoli2} possible viable
phenomenological frameworks to constrain the relative magnitude of
$\Lambda$ and $\delta$, while keeping  $M_h$ at the Fermi scale. For
instance, for $M_h=250$ GeV and $\Lambda\sim 10^{19}$ GeV one gets
$\delta\sim 10^{-5}$ eV and a ratio $\epsilon=\delta/M_h\sim 4\cdot
10^{-17}$ that might well represent the physical realization of a
formally infinitesimal quantity. If this were the right order of
magnitude, the non-Lorentz covariant density fluctuations of the
vacuum would start to show up from wavelengths larger than a few
millimeters up to infinity. These lengths are actually infinitely
larger than the Fermi scale but, nevertheless, have a physical
meaning.

Now, one might ask if there is any "first-principle" support for the
intuitive physical scenario sketched above. To this end, two
arguments were given in Ref.\cite{consoli} to motivate a possible
failure of the standard perturbative picture in the far infrared
region of the broken-symmetry phase. The first argument starts from
the separation of the full scalar field into a constant background
$\phi$ and a shifted fluctuation field $h(x)$ \BE \label{shift}
\Phi(x)=\phi + h(x) \EE Here, in order the separation to be
unambiguous, $\phi$ denotes the average field in a large four-volume
$\Omega$, i.e.
 \BE \phi={{1}\over{\Omega}}\int d^4x~ \Phi(x) \EE
 and the limit $\Omega \to \infty$ has to be taken at the end.
 In this way, the full functional measure can be expressed as
 \BE \label{measure}
 \int~[d\Phi(x)]= \int^{+\infty}_{-\infty}d\phi~\int [dh(x)] \EE
 where the functional integration on the right-hand side is over all
 quantum modes with four-momentum $p_\mu\neq 0$.

In the standard approach, one describes the shifted field $h(x)$ as
a purely massive field whose mass is related to the quadratic shape
of a non-convex semiclassical effective potential $V_{\rm NC}(\phi)$
at one of its absolute minima. This identification is based on
introducing, for any given $\phi$, the zero-momentum two-point
function $\Gamma_2(p=0)$,  i.e. the inverse zero-momentum connected
propagator $D^{-1}(p=0)$, through the relation \BE
D^{-1}(p=0)={{d^2V_{\rm NC} }\over{d\phi^2}} \EE and defining
$M^2_h$ from the value of $D^{-1}(p=0)$ at $\phi=\pm v$, the
absolute minima of $V_{\rm NC}(\phi)$.

In this series of steps one treats $\phi$, the zero-momentum mode of
the full scalar field $\Phi(x)$, as a purely classical object
without taking into account that, on the base of Eq.(\ref{measure}),
there should be one more functional integration over $d\phi$. In an
alternative approach, one can start from the generating functional
in the presence of a constant source $J$ \BE Z(J)=\
\int^{+\infty}_{-\infty}d\phi ~\exp[-\Omega(V_{\rm NC}(\phi)-J\phi)]
\EE where $V_{\rm NC}(\phi)$ is the non-convex effective potential
obtained, order by order in the loop expansion, after integrating
out all modes with $p_\mu\neq 0$.

Further, introducing the generating functional for connected Green's
functions $w(J)$ through \BE
  \Omega w(J)= \ln {{Z(J)}\over{Z(0)}} \EE
the vacuum field and the zero-momentum propagator are
  defined from the relations
  \BE \varphi(J)= {{dw}\over{dJ}} \EE and \BE G(p=0)= {{d^2w}\over{dJ^2}} \EE in the double limit $J \to 0$ and $\Omega
\to  \infty$.

Now, it is easy to show that in order to have SSB, i.e. \BE \lim_{J
\to 0^{\pm}}~\varphi(J) \neq 0 \EE the four-volume $\Omega$ has to
diverge in such a way that the dimensionless quantity $x=\Omega J v$
tends to a non-zero limit. However, even assuming that $x$ tends to
$\pm \infty$, so that the vacuum field $\varphi(J)$ tend to $ \pm
v$, still the zero-momentum inverse propagator $G^{-1}(p=0)$ does
not uniquely tend to $M^2_h$. In fact $G^{-1}(p=0)$ is the second
derivative of the exact effective potential that, as defined from
the Legendre transform of the generating functional, is not an
infinitely differentiable function in the infinite-volume limit
\cite{syma}. For this reason, in the broken-symmetry phase, there
are two solutions \BE \label{twosol}
G_a^{-1}(p=0)=M^2_h~~~~~~~~~~~~~~~~~~~~~~~~
 G_b^{-1}(p=0)=0 \EE
as if there were two different particles in the theory and not just
one.

In this sense, the situation of the broken phase is reminiscent of
what happens in superfluid $^4$He.  Also there the spectrum is
considered to arise from the combined effect of two different types
of excitations, phonons and rotons, whose separate energy spectra
match giving rise to a complicated pattern. Just following this
analogy, one can try to use quantum hydrodynamics \cite{epjc} to
obtain a physical interpretation of the parameter $M_h$  associated
with the massive part of the spectrum in terms of the energy-gap for
vortex formation in a superfluid medium possessing the same
constituents and the same density as the scalar condensate. In this
picture, where phonons would be physically cut off above those
infinitesimal momenta that correspond to the inverse-millimeter
range, the branch of the spectrum $\sim \sqrt { {\bf p}^2 + M^2_h}$
covers in practice the full range that is relevant for particle
physics.

The above derivation of two solutions for the zero-momentum
propagator is non-perturbative and independent of any diagrammatic
expansion. In Ref.\cite{consoli}, however, it was suggested that the
same conclusion is also obtained taking into account the effect of
the one-particle reducible, zero-momentum tadpole graphs. These
enter the diagrammatic expansion for the propagator in the presence
of a constant background field and can be considered a manifestation
of the quantum nature of the scalar condensate. These effects are
usually neglected in the standard procedure where, at the minima of
$V_{\rm NC}(\phi)$, one defines the connected propagator from a
Dyson sum of 1PI graphs alone. The reason is that the one-particle
reducible tadpole graphs are proportional to the one-point function
\BE J(\phi)= {{dV_{\rm NC} }\over{d\phi}} \EE that vanishes at
$\phi=\pm v$. However, these zero-momentum graphs are attached to
the other parts of the diagrams through zero-momentum propagators.
Therefore, ignoring their contribution at $\phi=\pm v$ (or including
their effect in a pure perturbative way with a form of the
zero-momentum propagator
 $G(p=0)={{1}\over{M^2_h}}+...$) is equivalent to assume the regularity
of $G(p=0)$ when $J\to 0$. Instead, if one relaxes this assumption
and looks for the general form of $G^{-1}(p=0)$, one finds again two
solutions as in Eq.(\ref{twosol}).

Addressing to Ref.\cite{consoli} for more details, the tacit
assumption at the base of the standard approach is better
illustrated with a very simple example. Consider the quadratic
equation \BE \label{full}
           f^{-1}(x)= 1 +x^2 -g^2 x^2 f(x)
\EE for $g^2 \ll 1$. The analogy with the problem of the
one-particle reducible zero-momentum tadpole graphs is established
when comparing $f(x)$ with $G(p=0)$, at a given value of $\phi$, and
the limit $x \to 0$ with the limit $J(\phi) \to 0$. Standard
perturbation theory is based on the iterative structure $f_{\rm
reg}=1/(1+x^2) + {\cal O}(g^2)$ that provides a class of solutions
that are regular for $x \to 0$ where $f_{\rm reg}(0)=1$. For this
class of solutions, that corresponds to the massive propagator as
defined from the one-particle irreducible graphs only, the third
term in the r.h.s. of Eq.(\ref{full}) vanishes identically for $x
\to 0$. On the other hand, Eq.(\ref{full}) has also a singular
solution $f_{\rm sing}\sim 1/g^2 x^2$ for $x\to 0$ and this
corresponds to a divergent zero-momentum propagator when $\phi \to
\pm v$. This can only be discovered by retaining the full
non-linearity of the problem where the zero-momentum propagators
joining to the vacuum sources are {\it not} approximated
perturbatively.

I realize that the various arguments given above cannot be
considered a "proof" that there is a gapless branch in the
excitation spectrum of the broken phase. They represent, however,
convergent indications that the standard perturbative derivation of
a massive spectrum, which is based on the quadratic part of the
shifted classical lagrangian after simply replacing $\phi=\pm v$ in
Eq.(\ref{shift}), may be too naive. A deeper understanding of the
ground state might be needed to obtain a proper description of the
far infrared region of the condensed phase. For instance, exploiting
the superfluid analogy, the methods of quantum hydrodynamics (see
e.g. Ref.\cite{jackiw}) might represent a natural alternative.

After this general introduction, I will provide in the rest of the
paper two more arguments that also indicate an unconventional
infrared behaviour. The first new argument, presented in Sect.2, is
based on the results of Ref.\cite{dario}. It uses the numerical
solution of the coupled RG equations for the effective potential and
field strength at various values of the infrared cutoff $k$.
Expressing the full scalar field as in Eq.(\ref{shift}), the
numerical results indicate that, in the broken phase, the
fluctuation field $h(x)$ cannot be represented in terms of purely
massive states in the $k \to 0$ limit.

The second new argument is based on a numerical simulation of the
broken phase in the Ising limit. If the shifted fluctuation field
were a purely massive field, the single-particle energy spectrum
should be reproduced by (the lattice version of) the standard
massive form $\sqrt { {\bf p}^2 + {\rm const.} }$ ~in the ${\bf p}
\to 0$ limit. At the same time, there should be no observable change
in the mass-gap increasing the linear lattice size above a typical
length scale associated with 7-8 correlation lengths. As discussed
in Sect.3 (details in the Tables at the end of the paper), both
expectations are not consistent with the lattice data.

Finally, Sect.4 will contain a summary and the conclusions.

\section{\normalsize{The RG equations for $V(\phi)$ and $Z(\phi)$. }}
\label{two}

The aim of Ref.\cite{dario} was to study the effective potential
$V(\phi)$ and the field strength $Z(\phi)$, as functions of the
background field $\phi$, at various values of the infrared cutoff.
This is a widely accepted technique where one starts from a bare
action defined at some ultraviolet cutoff $\Lambda$ and effectively
integrates out shells of quantum modes down to an infrared cutoff
$k$. This procedure generates a $k-$dependent effective action
$\Gamma_k[\Phi]$ that evolves into the full effective action
$\Gamma[\Phi]$ in the $k\to 0$ limit
\cite{weg,nicol,polch,wet1,mor1}. The $k-$dependence of
$\Gamma_k[\Phi]$ is determined by a differential functional flow
equation that is known in the literature in slightly different
forms.  In particular, with the flows discussed in detail in
Ref.\cite{wetreport} one starts form first principles and obtains a
class of functionals that interpolates between the classical bare
Euclidean action and the full effective action of the theory.
 However, some features, such as the basic convexity property
of the effective action for $k \to 0$
\cite{tetradis,litim2,alexander,zappala}, are independent of the
particular scheme.

In this approach, the relevant quantities are the $k-$dependent
effective potential $V_k(\phi)$ and field strength $Z_k(\phi)$,
which naturally appear in a derivative expansion of $\Gamma_k[\Phi]$
around a space-time constant configuration $\Phi(x)=\phi$. They are
governed by two coupled equations derived and discussed in
Refs.\cite{zappala,bonanno,litim12,litim11}.

Introducing dimensionless variables : $t={\rm ln}(\Lambda/k)$,
$x=k^{1-D/2}\phi$, $V(t,x)=k^{-D} V_k(\phi)$ and $Z(t,x)=Z_k(\phi)$
(where $D$ indicates the number of space-time dimensions) and
defining the first derivative of the effective potential
$f(x,t)=\partial_x V(t,x)$, these coupled equations can be expressed
in the form \cite{dario} ($f'=\partial_x f(t,x)$, $Z'=\partial_x
Z(t,x)$,...)
 \BE \label{fadim}
{{\partial f}\over{\partial t}} ={{(D+2)}\over {2}}f + {{(2-D)}\over
{2}}~x{{\partial f}\over{\partial x}} - {{1}\over{(4\pi)^{D/2}}}~
{{\partial }\over{\partial x}} e^{-f'/Z}  \EE \bea \label{zadim}
{{\partial Z}\over{\partial t}}&=& {{(2-D)}\over {2}}~x{{\partial
Z}\over{\partial x}}+
 {{1}\over{(4\pi)^{D/2}}}~
{{\partial }\over{\partial x}} \left (
{{Z'}\over{Z}}e^{-f'/Z}\right)\nonumber\\
-{{e^{-f'/Z}}\over{(4\pi)^{D/2} }}~
 &\Biggl(&{{f'(Z')^2}\over{Z^3}}+{{18D-D^2-20}\over{24}}
{{(Z')^2}\over{Z^2}} +{{(4-D)Z'f''}\over{6 Z^2}}- {{(f'')^2}\over{6
Z^2}}\Biggr ) \eea It is easy to show that these two coupled
equations can be transformed into the structure ($i,j$=1-3) \BE
P_{ij} {{\partial U_j}\over{\partial t}} + Q_i= {{\partial
R_i}\over{\partial x}} \EE where the components of the vector
$U_i(x,t)$ are the unknown functions of the problem and where
$P_{ij}$, $Q_i$ and $R_i$ can depend on $x,t, U_i, {{\partial
U_i}\over{\partial x}}$. In this way, the numerical solution was
obtained with the help of the NAG routines.

The analysis of Ref.\cite{dario} was focused on the quantum-field
theoretical case $D=4$ assuming standard boundary conditions at the
cutoff scale: i) a renormalizable form for the bare, broken-phase
potential \BE \label{bare} V_\Lambda(\phi)=-{{1}\over{2}}M^2\phi^2 +
\lambda \phi^4 \EE and ii) a unit normalization condition for the
derivative term in the bare action \BE Z_\Lambda(\phi)=1 \EE

It was also assumed a weak-coupling limit $\lambda=0.1$, fixing
$M=1$ and $\Lambda=10$. In this way, there is a well defined
hierarchy of scales where the infrared region corresponds to the
limit $k \ll M\ll \Lambda$.

The numerical results can be summarized as follows. For not too
small values of the infrared cutoff $k$, the effective potential
$V_k(\phi)$ remains a smooth, non-convex function of $\phi$ as in
the loop expansion. In this region of $k$ one also finds a field
strength $Z_k(\phi)\sim 1$ for all values of $\phi$.

However, a tiny scale $\delta\sim 0.15$ exists such that for $k
<\delta$ the effective potential $V_k(\phi)$ starts to flatten in an
inner region of $|\phi|$ while still matching with an outer,
asymptotic shape of the type expected in perturbation theory. The
flattening in the inner $|\phi|$-region, while reproducing the
expected convexity property of the effective potential, does not
correspond to a smooth behaviour.

For such small values of $k$ there are large departures of
$Z_k(\phi)$ from unity in the inner $|\phi|-$region with a strong
peaking at the end point $|\hat{\phi}|=|\hat{\phi}(k)|$ of the
flattening region. On the base of the general convexification
property, the $k \to 0$ limit of such end point, $\hat{\phi}(0)$,
coincides with one of the minima $\pm v$ of a suitable
semiclassical, non-convex effective potential and is usually taken
as the physical realization of the broken phase.

To interpret these results, let us start from the usual point of
view where SSB is described in terms of a classical potential with
perturbative quantum corrections. These corrections, with the choice
of the bare parameters of Ref.\cite{dario}, are typically small for
all quantities. In particular $Z$, in perturbation theory, is a
non-leading quantity since its one-loop correction is ultraviolet
finite. Therefore, perturbation theory predicts tiny deviations of
$Z$ from unity, $ \sim 10^{-2}$. This is also in agreement with the
assumed exact "triviality" of the theory \cite{triviality} that
requires $Z \to 1$ in the continuum limit.

This prediction fits well with the profile of $Z_k(\phi)$ for not
too small values of the infrared cutoff. However, for $k <\delta$,
there are large deviations from unity with the mentioned strong
peaking phenomenon.

If we express the full scalar field $\Phi(x)$ as in
Eq.(\ref{shift}), the above results indicate that the higher
frequency components of the fluctuation field $h(x)$, those with
4-momentum $p_\mu$ such that $\delta \leq |p|\leq \Lambda$,
represent genuine quantum corrections for all values of the
background field $\phi$ in agreement with their perturbative
representation as weakly coupled massive states.

On the other hand, the components with a 4-momentum $p_\mu$ such
that $ |p| \leq \delta$,  are non-perturbative for values of the
background field in the range $0 \leq \phi \leq \hat{\phi}(|p|)$. In
particular, the very low-frequency modes with $|p| \to 0$ behave
non-perturbatively for all values of the background in the full
range $0 \leq |\phi| \leq v$ and thus cannot be represented as
standard massive states. In fact, non-perturbative infrared
phenomena cannot occur in a genuine massive theory.

Notice that the unexpected effects show up in connection with the
convexification process, precisely as discussed in
Ref.\cite{consoli} and reviewed in the Introduction. In this sense,
one can say that the convexification process is induced by the
infinitely long-wavelength modes that, so to speak, "live" in the
full region $0\leq |\phi| \leq v$.

Notice also that, by itself, the existence of a non-perturbative
infrared sector in a region $0\leq |p| \leq \delta$ might not be in
contradiction with the assumed exact "triviality" property of the
theory if, in the continuum limit, the infrared scale $\delta$
vanishes in units of the physical parameter $M_h$ associated with
the massive part of the spectrum. This means to establish a
hierarchy of scales $\delta \ll M_h \ll \Lambda$ such that
${{\delta}\over{M_h}}\to 0$ when ${{M_h}\over{\Lambda}} \to 0$, as
discussed in the Introduction.

If this happens, the region $0\leq |p| \leq \delta$ would just
shrink to the zero-measure set $p_\mu=0$, for the continuum theory
where $M_h$ sets the unit mass scale, thus recovering the exact
Lorentz covariance of the energy spectrum since the point $p_\mu=0$
forms a Lorentz-invariant subset. In this limit, the RG function
$Z_k(\phi)$ would become a step function which is unity for all
finite values of $k$ (and $\phi$) and is only singular for $k=0$ in
the range $0 \leq |\phi| \leq v$. In this way, one is left with a
massive, free-field theory for all non-zero values of the momentum,
and the only remnant of the non-trivial infrared sector is the
singular re-scaling of $\phi$ (the projection of the full scalar
field $\Phi(x)$  onto $p_\mu=0$).

\section{\normalsize{Lattice simulation of the broken phase in the Ising limit}}
\label{two}

In this section I will present the results of a numerical simulation
of the theory. The main goal of the numerical experiment is to check
the standard assumption that for ${\bf k } \to 0$ the excitation
spectrum of the broken phase is the same as in a weakly coupled
massive theory. Quite independently of the theoretical arguments
given in the Introduction and in Sect.2, the idea of deviations from
the simple massive behaviour of perturbation theory is supported by
the lattice results of Ref.\cite{Cea2}. There, differently from what
happens in the symmetric phase, the connected scalar propagator was
found to deviate significantly from (the lattice version of) the
massive single-particle form $1/(p^2+{\rm const})$ for $p_\mu \to
0$. In particular, looking at Figs. 7, 8 and 9 of
Ref.\cite{stevensonnp}, one can clearly see that, approaching the
continuum limit of the lattice theory, these deviations become more
and more pronounced but also confined to a smaller and smaller
region of momenta near $p_\mu=0$.

As in Ref.\cite{Cea2}, the test has been performed in the Ising
limit that traditionally has been chosen as a convenient laboratory
for the numerical analysis of the theory. In this limit, a
one-component $\Phi^4_4$ theory becomes governed by the lattice
action
\begin{equation}
\label{ising} S_{\text{Ising}} = -\kappa \sum_x\sum_{\mu} \left[ \phi(x+\hat
e_{\mu})\phi(x) + \phi(x-\hat e_{\mu})\phi(x) \right]
\end{equation}
where $\phi(x)$ takes only the values $\pm 1$. The broken-symmetry
phase corresponds to values of the hopping parameter
$\kappa>\kappa_c$ with $\kappa_c\sim 0.07484$ \cite{critical}.

Observables include the bare magnetization:
\begin{equation}
\label{baremagn}
 v_B=\langle |\phi| \rangle \quad , \quad \phi \equiv \sum_x
\phi(x)/L^4
\end{equation}
(where $\phi$  is the average field for each lattice configuration) and the
bare zero-momentum susceptibility:
\begin{equation}
\label{chi}
 \chi_{\text{latt}}=L^4 \left[ \left\langle |\phi|^2
\right\rangle - \left\langle |\phi| \right\rangle^2 \right] .
\end{equation}
The equivalence of these two definitions with other standard lattice
definitions that are found in the literature has been discussed in
Ref.~\cite{balog}.

Let us now consider the mass gap of the theory, say
$m_{\text{TS}}({\bf{k}}=0)$, that is traditionally extracted at zero
3-momentum from the exponential decay (TS=`Time Slice') of the
connected two-point correlator. This is a well known strategy (see
for instance Ref.~\cite{montmunster}) that, however, for the sake of
clarity it might be useful to review in some detail.

The Fourier transform of the connected two-point correlator can be
expressed as
\begin{equation}
\label{corr} C_1(t,0; {\bf k})\equiv \langle S_c(t;{\bf
k})S_c(0;{\bf k})+ S_s(t;{\bf k})S_s(0;{\bf k}) \rangle _{\rm conn}
\end{equation}
where
\begin{equation}
\label{cos} S_c(t; {\bf k})\equiv \frac{1}{L^3} \sum _{ { \bf x} }
\phi({\bf x}, t) \cos ({\bf k} \cdot {\bf x}) ,
\end{equation}
\begin{equation}
\label{sin} S_s(t;{ \bf k})\equiv \frac{1}{L^3} \sum _ {{\bf x}}
\phi({\bf x}, t) \sin ({\bf k} \cdot {\bf x}) .
\end{equation}
Here, $0\leq t\leq L_t$ is the Euclidean time; ${\bf x}$ is the
spatial part of the site 4-vector $x^{\mu}$; ${\bf k}$ is the
lattice 3-momentum ${\bf k}=(2\pi/L) (n_x,n_y,n_z$), with
$(n_x,n_y,n_z)$ non-negative integers; and $\langle ...\rangle_{\rm
conn}$ denotes the connected expectation value with respect to the
lattice action, Eq.~(\ref{ising}).

To obtain the time-slice mass, one starts from the general
expression (see Ref.~\cite{montmunster})
\begin{equation}
\label{fitcor21} C_1(t,0;{\bf k})= \sum_\alpha |A_\alpha|^2 e^{
-E_\alpha ({\bf{k}})t }
\end{equation}
where the sum is over the eigenstates of the lattice Hamiltonian
corresponding to the given value of ${\bf{k}}$. In perturbation
theory, these are represented in the Fock space as massive
$\alpha-$particle states coupled to total momentum ${\bf{k}}$ and
one predicts $E_1({\bf{k}}) \leq E_2({\bf{k}}) \leq ...$. Therefore,
using the lattice dispersion relation
\begin{equation}
\label{disp3} m^2_{\rm TS}({\bf k}) = ~2 (\cosh E_1({\bf{k}})  -1)~~
-~~2 \sum ^{3} _{\mu=1}~(1-\cos k_\mu) \,
\end{equation}
one can extract the mass from the single-particle energy
$E_1({\bf{k}})$.

Now, in an interacting theory, a model-independent approach to the
energy spectrum would require to extract the leading term
$E_1({\bf{k}})$ from the exponential decay at asymptotic $t$.
However, in a real numerical simulation the vanishing of the
correlator can hardly be studied for {\it asymptotic} times with
enough statistics. Thus, in order to extract the leading
single-particle energy, one is forced to introduce some
model-dependent assumptions, such as the shape of $E_1({\bf{k}})$,
its relation with the higher excitations $E_2({\bf{k}}),
E_3({\bf{k}}),..$ and so on.

On the other hand, in a weakly coupled theory (as in a "trivial"
theory close to the continuum limit), there is a simple strategy to
extract the single particle energy that has the advantage of being
free of uncontrolled theoretical assumptions. It can be applied once
the ratio
\begin{equation}
\label{ratio} R = \frac{|A_1|^2 }{ \sum_\alpha |A_\alpha|^2} \leq 1
\end{equation}
is expected to be very close to unity. In this regime, by further
noticing that the residual corrections to a single-particle
correlator are of order \BE (1-R)e^{-\Delta_n({\bf{k}})t} \EE  with
$\Delta_n({\bf{k}})=E_n({\bf{k}})-E_1({\bf{k}})>0$ and
n=2,3,$\dots$, and thus are suppressed by additional phase-space
factors, one can start parameterizing the full correlator (with
periodic boundary conditions) in terms of an effective
single-particle spectrum
\begin{equation}
\label{fitcor2} C_1(t,0;{\bf k})= A \, [ \, \exp(-E({\bf{k}})
t)+\exp(-E({\bf{k}})(L_t-t)) \, ] \,
\end{equation}
Whenever higher excited states were really needed to describe the
lattice data, one should detect appreciable differences, both in the
value of the normalized chi-square and in the fitted value of
$E({\bf{k}})$,  by simply varying the range of $t$ where the fit is
performed. This means to look for the stability of the results
finding a characteristic `plateau' $E^{(p)}({\bf{k}})$ (p=`plateau')
for the fitted energy where the different indications converge and
that can be used to safely extract the time-slice mass.

After these preliminaries, let us now consider the numerical
results. The simulations were performed with a numerical code
written by Paolo Cea and Leonardo Cosmai. It employs the
Swendsen-Wang cluster algorithm~\cite{Swendsen}, using the cluster
improved estimator~\cite{frick} to compute the time slice
correlations. The statistical analysis is performed using the
jackknife method \cite{efron}.

To illustrate with an example the determination of the energy
plateau, let us first look  at the symmetric phase for
$\kappa=0.074$. The raw data for the connected correlator are
reported in Tables 1, 2 and 3 for three values of the lattice
3-momentum ${\bf{k}}^2=0$, ${\bf{k}}^2=0.375$ and ${\bf{k}}^2=0.922$
and two lattice sizes. I also show the quality of the fits and the
resulting time-slice mass in various ranges of $t$. As one can see,
after simply skipping the first one or two time slices, one gets
very good stability with mass values that are independent of the
spatial momentum and in excellent agreement with each other. This
confirms to a very high precision that, in the symmetric phase, the
single-particle energy spectrum of the theory is well reproduced by
a standard massive spectrum.

Let us now consider the broken phase. This was simulated choosing
the value of the hopping parameter $\kappa=0.076$. At this value of
$\kappa$, in fact, published data by Balog et al. \cite{balog} from
a $20^4$ lattice provide the estimate $m_{\rm TS}(0)=0.392(1)$. A
check that this is a real mass gap can be obtained in two ways.
First, as done for the symmetric phase, one can study the stability
of $ m_{\rm TS}({\bf k})$ at various values of the spatial momentum.
Second, one can check the dependence on the lattice size. In fact,
if $m_{\rm TS}(0)$=0.392(1) were a real mass gap, a $20^4$ lattice
contains already $\sim 8$ correlation lengths and, thus, there
should be no significant change by further increasing the lattice
size.

The simulations were first performed on $20^4$ lattices at different
values of ${\bf{k}}^2$. In Table 4, I report the results of a
simulation of 6 Msweeps for ${\bf{k}}=0$.  As one can check, the
magnetization and the susceptibility are in excellent agreement with
those reported by Balog et al.  for the same lattice size (see the
corresponding entries for $\kappa=0.076$ in Table 3 of
Ref.\cite{balog}).

Concerning the time-slice mass, the value $m_{\rm TS}(0)=$0.3920(24)
is also in excellent agreement with the result of Ref.\cite{balog}.
In particular, using all data in the t-range 1-19 the fit with Eq.
(\ref{fitcor2}) goes through all central values with a {\it total}
chi-square which is less than 0.1. This shows that, for
${\bf{k}}=0$, the deviations of the correlator data from a pure
single exponential are confined to the first time slice.

The results for ${\bf{k}}\neq 0$ are reported in Tables 5-9. The
number of sweeps is such to provide statistical errors of the
time-slice mass that are approximately equal to those reported in
Table 4 for ${\bf{k}}=0$. As one can see, differently from the
symmetric phase, there is a distinct dependence of $ m_{\rm TS}({\bf
k})$ on the spatial momentum (that was already observed in
Ref.\cite{Cea2}).

Let us first compare with the data in Tables 5 and 6. These refer to
the lowest two momenta of a $20^4$ lattice. Again, as for
${\bf{k}}=0$, the deviations from a pure single exponential are
limited to the first time slice since the fit in the full t-range
1-19 with Eq. (\ref{fitcor2}) goes through all central values to a
very high precision. Using the results of the fit in this range one
obtains time-slice masses $ m_{\rm TS}({\bf k})=$ 0.3992(21) and
0.4039(21) for the integer assignments of the 3-momentum (1,0,0) and
(1,1,0) respectively. These values are not consistent within 3-5
$\sigma$ with the corresponding value 0.3920(24) reported in Table
4. Therefore, the lattice data of the  broken phase are not well
reproduced by the standard massive form in the  ${\bf{k}}\to 0$
limit.

Let us now consider the data at the higher spatial momenta reported
in Tables 7-9. From the substantial equivalence of the fits in the
ranges 2-18 and 3-17, one can deduce that, in this momentum range,
the deviations from a pure single exponential affect now the first
two time slices. However, notice the difference with the symmetric
phase. From Table 3, using the fit results in the t-range 2-18, one
obtains $ m_{\rm TS}({\bf k})=$0.2133(33) in very good agreement
with the mass values reported in Table 1. From Tables 7-9, on the
other hand, one obtains the corresponding entries $ m_{\rm TS}({\bf
k})=$ 0.4122(15), 0.4127(18), 0.4129(20) that are not consistent
within 7-8 $\sigma$ with the mass value 0.3920(24) of Table 4. Using
the fit results in the t-range 3-17, statistical errors become
larger but the discrepancy remains at the level of 3-5 $\sigma$.

On the basis of the previous results for ${\bf{k}}\neq 0$, it should
be clear that interpreting the value $m_{\rm TS}(0)=0.392(1)$ as a
real mass gap is not so obvious. Actually, if one looks for
stability with respect to changes of the spatial momentum, it is
only at {\it higher} momenta that one gets consistency with the
standard concept of "mass" as a ${\bf k}$-independent quantity. In
fact, looking at Tables 7-9, if one requires both a good-quality fit
and a substantial independence of the chi-square per degree of
freedom on the sample of data (as it happens for the fits 2-18 and
3-17) one obtains a remarkable momentum independence for a mass
value $\sim 0.412(3)$. This confirms the basic idea of the
Introduction, namely that the shifted fluctuation field is not
purely massive and that the end point of the spectrum at ${\bf{k}}=0$ cannot be used to extract the "mass".

The idea that $m_{\rm TS}(0)$ might not be a real, physical mass gap
is also confirmed by its {\it volume dependence}. To this end, I
report, in Tables 10-13 the raw correlator data from $32^4$
lattices. The total statistics is 6 Msweeps as for the simulation at
${\bf{k}}=0$ on the $20^4$ lattice reported in Table 4. To shorten
the total time of the numerical experiment, however, the full
statistics, for the $32^4$ case, was collected using different
computers. In the $32^4$ case, I only report data for which the S/N
ratio is larger than unity.

Notice that the magnetization and the susceptibility are completely
consistent with those measured on the $20^4$ lattice. On the other
hand, looking at the results of the global fit to the correlator
data, one finds a serious discrepancy in the value of the mass gap.
In fact both from Table 4 and Table 14, there are clean indications
for the existence of an energy plateau. However the two values are
substantially different. In particular, the $32^4$ value
$E^{(p)}(0)\sim 0.364(5)$ gives a time slice mass $m_{\text{TS}}(0)=0.366(5)$ that is not consistent within 5 $\sigma$ with the value
0.392(1) reported by Balog et al. for the $20^4$ lattice.

As an additional check, a fit with 2 exponentials was also performed
along the lines indicated in Ref.\cite{balog2}. In this case,
Eq.(\ref{fitcor2}) is replaced by a constrained 2-mass formula with
$E_2(0)=2E_1(0)$. The results of this other type of fit are shown in
Tables 15 and 16. The entries that are unchanged correspond to fits
where the normalization of the 2-particle term is set to zero by the
fit routine. For other entries, the statistical errors become much
larger. However, the discrepancy between the mass value from the
$20^4$ lattice and that obtained from the $32^4$ lattice remains.

Notice that, differently from $m_{\text{TS}}(0)$ the susceptibility
$\chi_{\rm latt}$ remains almost unchanged when increasing the
lattice size. Since the susceptibility is nothing but the
zero-four-momentum connected propagator, one can express the
numerical stability of the result in the form $ \chi_{\rm
latt}=G_a(p=0) $ thus relating $\chi_{\rm latt}$ to the a-type,
massive solution for the zero-momentum propagator in
Eq.(\ref{twosol}). On the other hand, the lattice results suggest
also that, if there were a gapless branch of the shifted field, the
lattice definition of the b-type of solution might be identified
through the relation $m^2_{\rm TS}(0)=G_b^{-1}(p=0)$. To prove this
conjecture, one should check that increasing the lattice size one
gets smaller and smaller values of the mass gap. To get a clean
indication, for the same value $\kappa=0.076$, this will require the
non-trivial task to compute the connected two-point correlator, with
the same Swendsen-Wang cluster algorithm and a statistics of $\sim6$
Msweeps, on a substantially larger lattice, say on a $48^4$.

Before ending this section, I observe that from the results of
Sect.2 the non-perturbative infrared sector of the broken phase
emerges as a {\it threshold} phenomenon starting at a momentum scale
$\delta \ll M_h$. As anticipated, a massive free-field limit for the
continuum theory implies a hierarchical structure of scales where
the ratio $\epsilon={{\delta}\over{M_h}}\to 0$ when $t
={{\Lambda}\over{M_h}}\to \infty$. Therefore, the discrepancy in the
values of the mass gap, that is already observed at $\kappa=0.076$
by simply increasing $L$ from 20 to 32, would show up on larger and
larger lattice sizes $L > 1/\delta$ for $\kappa$ approaching the
critical value $\kappa_c\sim 0.07484$ \cite{critical}.

For instance assuming a hierarchical relation of the type in
Eq.(\ref{golden}), the minimum lattice size would be $L >
\Lambda/M^2_h$. Thus, using the relation \cite{brahm} $\Lambda\sim
{{1.5\pi}\over{a}}$, to express the ultraviolet cutoff of a
$\lambda\Phi^4_4$ theory in units of the inverse lattice spacing,
one should expect \BE {{L}\over{a}} > {{1.5\pi}\over{(M_ha)^2}} \EE
This gives ${{L}\over{a}} >$ 29, 52, 118, 471 for $M_h a$=0.4, 0.3,
0.2 , 0.1 respectively. In this sense, the range close to
$\kappa=0.076$, where $M_h a\sim $0.4, is uniquely singled out since
it lies in the scaling region but, at the same time, the required
lattice sizes are not too large. For the same reason, the stability
of a mass gap such as 0.205 when increasing the lattice size from
${{L}\over{a}}=36$ to ${{L}\over{a}}=48$ (see Table 4 of
Ref.\cite{balog}) might be consistent with the results presented
here. In this case, in fact, to detect the same discrepancy found
for $M_h a\sim 0.4$ a minimum value ${{L}\over{a}}\sim 112$ might be
needed.

\section{\normalsize{Summary and conclusions}}
\label{three}

In this paper, following the original idea of Ref.\cite{consoli}, I
have presented additional evidences that the conventional
perturbative picture of SSB might miss an important infrared
phenomenon: the standard singlet Higgs boson might not be a purely
massive field. In fact, as reviewed in the Introduction, there are
arguments to expect a non-perturbative infrared sector corresponding
to the long-wavelength excitations of the scalar condensate.

On a more formal ground, as a consequence of the convexification
process of the effective potential, the inverse zero-momentum
connected propagator should be considered a two-valued function
that, besides the standard massive solution $G_a^{-1}(p=0)=M^2_h$,
includes the value $G_b^{-1}(p=0)=0$ as in a gapless theory. As
shown in Ref.\cite{dario} and illustrated in Sect.2, the
convexification process is a threshold phenomenon that starts when
the infrared cutoff $k$ is below a tiny scale $\delta$. In this
regime, where the effective potential $V_k(\phi)$ deviates from the
smooth semiclassical form of perturbation theory, the field strength
$Z_k(\phi)$ starts to exhibit large deviations from unity with a
strong peaking phenomenon that extends to the boundary of the
flatness region in the limit $k\to 0$. Such a behaviour could hardly
be explained if the infrared region were that of a simple massive
theory.

In Sect. 3 (and in the Tables at the end of the paper) I have
produced detailed numerical results from a lattice simulation of the
broken phase in the 4D Ising limit of the theory. They point to the
following conclusions:

~~~i)  differently from the symmetric phase, the single-particle
energy spectrum is not well reproduced by (the lattice version of)
the standard massive form $\sqrt { {\bf p}^2 + {\rm const.} }$ in
the limit ${\bf p} \to 0$ and a $|{\bf p}|$-independent mass
parameter is only found at higher momenta

~~ii) for the value $\kappa=0.076$, the mass gap reported by Balog
et al.\cite{balog} on a $20^4$ lattice is
$m_{\text{TS}}(0)=0.392(1)$. On the other hand increasing the
lattice size up to $32^4$ the data provide $m_{\text{TS}}(0)=0.366(5)$, contrary to the expectation that there should be no
significant change. At the same time, since the susceptibility is
practically unchanged, this might represent one more evidence for
the existence of two solutions for the zero-momentum propagator and
for the subtle nature of the zero-momentum limit in the
broken-symmetry phase.

For this reason, the observed volume dependence of
$m_{\text{TS}}(0)$, with its possible interpretation in terms of
(`non-Goldstone') collective excitations of the scalar condensate,
poses interesting questions and would deserve to be checked by other
groups with new systematic investigations of the $L\to \infty$ limit
of the lattice theory. To this end, one should also take into
account that the deviations from a simple massive spectrum $\sqrt {
{\bf p}^2 + M^2_h}$ might be confined to a region of momenta $|{\bf
p}| < \delta$ where $\delta$ vanishes in units of $M_h$ in the
infinite-cutoff limit. This means that, approaching the continuum
limit of the lattice theory, the volume dependence of the mass gap
might require larger and larger lattice sizes before to show up.
\vskip 25pt \centerline{{\bf Acknowledgements}} \vskip 10 pt I wish
to thank Paolo Cea and Leonardo Cosmai for giving me all necessary
ingredients for the lattice simulations. I also thank Dario
Zappal\`a for many useful discussions.

\vfill\eject


\begin{thebibliography}{10}

\bibitem{thooft}
G. 't Hooft, In Search of the Ultimate Building Blocks, Cambridge
University Press, 1997, p.70.

\bibitem{mech}
M. Consoli and P.M. Stevenson, Int. J. Mod. Phys. {\bf A15} (2000)
133, hep-ph/9905427.



\bibitem{consoli}
M. Consoli, Phys. Lett. {\bf B512} (2001) 335; Phys. Rev. {\bf D65}
(2002) 105217.

\bibitem{stevenson2001}
P. M. Stevenson, in Proceedings of the Conference on Violations of
CPT and Lorentz invariance, Indiana 2001 (World Scientific,
Singapore 2002).

\bibitem{voloreport}
G. E. Volovik, Phys. Rep. {\bf 351} (2001) 195.

\bibitem{consoli2}
M. Consoli, Phys. Lett. {\bf B541} (2002) 307.

\bibitem{volo}
G. E. Volovik, JETP Lett.{\bf 73} (2001) 162.

\bibitem{grifols}
F. Ferrer and J. Grifols, Phys. Rev. {\bf D63} (2001) 025020.

\bibitem{syma}
K. Symanzik, Commun. Math. Phys. {\bf 16} (1970) 48.

\bibitem{epjc}
M. Consoli and E. Costanzo, Eur. Phys. J. {\bf C33} (2004) 297.



\bibitem{jackiw}
R. Jackiw, S. -Y. Pi and A. Polychronakos, Ann. Phys. {\bf 301}
(2002) 157.

\bibitem{dario}
M. Consoli and D. Zappal\`a, arXiv:hep-th/0606010, to appear in
Phys. Lett. {\bf B}.

\bibitem{weg} F. J. Wegner and A. Houghton, Phys. Rev {\bf A8} (1973) 401.

\bibitem{nicol}
T.S. Chang, D. D. Vvedensky and  J.F. Nicoll, Phys. Rep. {\bf 217}
(1992) 280.

\bibitem{polch} J. Polchinski, Nucl. Phys. {\bf B 231} (1984) 269.

\bibitem{wet1}
C. Wetterich, Nucl. Phys. {\bf B352}, 529 (1991); Phys Lett. {\bf
B301} (1993) 90.

\bibitem{mor1} T. Morris, Int. J. Mod. Phys. {\bf A 9} (1994) 2411;
Phys. Lett. {\bf B329} (1994) 241.

\bibitem{wetreport} J. Berges, N. Tetradis and  C. Wetterich,
Phys.Rep {\bf 363} (2002) 223.

\bibitem{tetradis}
K.-I. Aoki, A. Horikoshi, M. Taniguchi and H. Terao, The Exact
Renormalization Group, Proceedings of the Workshop on the Exact
Renormalization Group, Faro, Portugal, September 10-12 1998, (World
Scientific 1999), arXiv:hep-th/9812050; A. S. Kapoyannis and N.
Tetradis, Phys. Lett. {\bf A 276} (2000) 225.

\bibitem{litim2} D. F. Litim, J. M. Pawlowski and L. Vergara, Convexity of
the effective action from functional flows , arXiv:hep-th/0602140.

\bibitem{alexander} J. Alexandre, V. Branchina and J. Polonyi, Phys.
Lett. {\bf 445} (1999) 351.

\bibitem{zappala} D. Zappal\`a, Phys. Lett. {\bf A290} (2001) 35.


\bibitem{bonanno} A. Bonanno and  D. Zappal\`a, Phys. Lett.  {\bf B504} (2001)
181; M.Mazza and  D. Zappal\`a, Phys. Rev. {\bf D64} (2001) 105013.

\bibitem{litim12} D.F. Litim and  J. M. Pawlowski,
Phys. Lett. {\bf  B546} (2002) 279; Phys. Rev. {\bf D66} (2002)
025030.

\bibitem{litim11} D.F. Litim and  J. M. Pawlowski,
Phys. Lett. {\bf  B516} (2001) 197; Phys. Rev. {\bf D65} (2002)
081701.


\bibitem{triviality}
For a review of the rigorous results, see R. Fern\'andez, J.
Fr\"ohlich, and A. D. Sokal, Random Walks, Critical Phenomena and
Triviality in Quantum Field Theory  (Springer-Verlag Berlin
Heidelberg 1992).


\bibitem{Cea2}
P. Cea. M. Consoli, L. Cosmai and P. M. Stevenson, Mod. Phys. Lett.
{\bf A14} (1999) 1673, hep-lat/9902020.

\bibitem{stevensonnp}
P. M. Stevenson, Nucl. Phys. {\bf B729} (2005) 542.

\bibitem{critical}
D.S. Gaunt, M.F. Sykes and S. McKenzie, J. Phys. {\bf A12} (1979)
871; D. Stauffer and J. Adler, Int. J. Mod. Phys. {\bf C8} (1997)
263.

\bibitem{balog} J. Balog et~al., Nucl. Phys. {\bf B714} (2005) 256,
hep-lat/0412015.

\bibitem{montmunster}
I. Montvay and G. M{\"u}nster, Quantum Fields on a Lattice,
Cambridge University Press, 1994.


\bibitem{Swendsen}
R.H. Swendsen and J.S. Wang,
\newblock Phys. Rev. Lett. {\bf 58} (1987) 86.

\bibitem{frick}
C. Frick, K. Jansen, and P. Seuferling, Phys. Rev. Lett. {\bf 63}
(1989) 2613.

\bibitem{efron}
B. Efron, Jackknife, the Bootstrap and other resampling plans (SIAM,
1982); B. A. Beg and A. H. Billoire, Phys. Rev. {\bf D 40} (1989)
550.


\bibitem{balog2}
J. Balog, F. Niedermayer and P. Weisz, hep-lat/0601016.

\bibitem{brahm}
D. E. Brahm, arXiv:hep-lat/9403021.


\end{thebibliography}

\vfill\eject


\begin{table}
\begin{tabular}{ccc}
$t$  & $C_1(t,0;{\mathbf k}=0)$ &  ${\text{statistical error}}$ \\
\hline
   0     &0.46445456707497D-03   &0.46037664989336D-06 \\
   1     &0.37571910789012D-03   &0.46116594635711D-06 \\
   2     &0.30382577283847D-03   &0.45018527889238D-06 \\
   3     &0.24582066021523D-03   &0.43438332218099D-06 \\
   4     &0.19905074016424D-03   &0.41560747498003D-06 \\
   5     &0.16136544145195D-03   &0.39542065913881D-06 \\
   6     &0.13104967814461D-03   &0.37531332727689D-06 \\
   7     &0.10671304049363D-03   &0.35557753750683D-06 \\
   8     &0.87238915133100D-04   &0.33706071111197D-06 \\
   9     &0.71738763750074D-04   &0.32019883688822D-06 \\
  10     &0.59502097899827D-04   &0.30574726422856D-06 \\
  11     &0.49970839560653D-04   &0.29433390224131D-06 \\
  12     &0.42700968213871D-04   &0.29772329849853D-06 \\
  13     &0.37366513141293D-04   &0.29930693571779D-06 \\
  14     &0.33719361396463D-04   &0.29930065059117D-06 \\
  15     &0.31592799783943D-04   &0.29950600189874D-06 \\
  16     &0.30893472344528D-04   &0.30005725246790D-06 \\
  17     &0.31592799783943D-04   &0.29950600189874D-06 \\
  18     &0.33719361396463D-04   &0.29930065059117D-06 \\
  19     &0.37366513141293D-04   &0.29930693571779D-06 \\
  20     &0.42700968213871D-04   &0.29772329849853D-06 \\
  21     &0.49970839560653D-04   &0.29433390224131D-06 \\
  22     &0.59502097899827D-04   &0.30574726422856D-06 \\
  23     &0.71738763750074D-04   &0.32019883688822D-06 \\
  24     &0.87238915133100D-04   &0.33706071111197D-06 \\
  25     &0.10671304049363D-03   &0.35557753750683D-06 \\
  26     &0.13104967814461D-03   &0.37531332727689D-06 \\
  27     &0.16136544145195D-03   &0.39542065913881D-06 \\
  28     &0.19905074016424D-03   &0.41560747498003D-06 \\
  29     &0.24582066021523D-03   &0.43438332218099D-06 \\
  30     &0.30382577283847D-03   &0.45018527889238D-06 \\
  31     &0.37571910789012D-03   &0.46116594635711D-06 \\
  32     &0.46445456707497D-03   &0.46037664989336D-06 \\
\end{tabular}
\\[1.0cm]
\begin{tabular}{ccccc}
$t-range$  & $\chi^2$/d.o.f. & ${\bf{k}}^2$ &  $E({\bf{k}})$ &
$m_{\text{TS}}({\bf{k}})$  \\
\hline
0-32    &0.072     & 0.000  & 0.2129 (2) &  0.2133 (2)   \\
1-31    &0.075     & 0.000  & 0.2128 (2) &  0.2133 (2)    \\
2-30    &0.052     & 0.000  & 0.2128 (3) &  0.2132 (3)    \\
3-29    &0.035     & 0.000  & 0.2127 (3) &  0.2131 (3)     \\
\end{tabular}
\caption{The raw correlator data obtained for a spatial momentum
${\bf{k}}=0$, on a $32^4$ lattice, for the value $\kappa=0.074$ in
the symmetric phase. The statistics is 135Ksweeps. I also show the
quality of the fits, the effective single-particle energy and the
time-slice mass obtained by using Eqs.(\ref{fitcor2}) and
(\ref{disp3}) in various ranges of $t$.} \label{Table1}
\end{table}

\begin{table}
\begin{tabular}{ccc}
$t$  & $C_1(t,0;{\mathbf k})$ &  ${\text{statistical error}}$ \\
\hline
   0    &0.14623108565057D-03  &0.14165730212821D-07 \\
   1    &0.77214443895274D-04  &0.13288355211454D-07 \\
   2    &0.40796525737581D-04  &0.11455026260932D-07 \\
   3    &0.21558338760062D-04  &0.95795734647824D-08 \\
   4    &0.11389647365225D-04  &0.81678791276023D-08 \\
   5    &0.60169117411735D-05  &0.78549959762898D-08 \\
   6    &0.31785227732011D-05  &0.64601886340349D-08 \\
   7    &0.16760427054163D-05  &0.57641678003130D-08 \\
   8    &0.88228124131541D-06  &0.53085501360843D-08 \\
   9    &0.46368972631045D-06  &0.46940187418747D-08 \\
  10    &0.24302326620048D-06  &0.42367563032742D-08 \\
  11    &0.12847759060431D-06  &0.38866243669447D-08 \\
  12    &0.68535903339602D-07  &0.37374075184433D-08 \\
  13    &0.37768265153790D-07  &0.37369789004914D-08 \\
  14    &0.22751593190996D-07  &0.39349497056091D-08 \\
  15    &0.15386378604173D-07  &0.42589381877184D-08 \\
  16    &0.13380756871513D-07  &0.44421054740855D-08 \\
  17    &0.15386378604173D-07  &0.42589381877184D-08 \\
  18    &0.22751593190996D-07  &0.39349497056091D-08 \\
  19    &0.37768265153790D-07  &0.37369789004914D-08 \\
  20    &0.68535903339602D-07  &0.37374075184433D-08 \\
  21    &0.12847759060431D-06  &0.38866243669447D-08 \\
  22    &0.24302326620048D-06  &0.42367563032742D-08 \\
  23    &0.46368972631045D-06  &0.46940187418747D-08 \\
  24    &0.88228124131541D-06  &0.53085501360843D-08 \\
  25    &0.16760427054163D-05  &0.57641678003130D-08 \\
  26    &0.31785227732011D-05  &0.64601886340349D-08 \\
  27    &0.60169117411735D-05  &0.78549959762898D-08 \\
  28    &0.11389647365225D-04  &0.81678791276023D-08 \\
  29    &0.21558338760062D-04  &0.95795734647824D-08 \\
  30    &0.40796525737581D-04  &0.11455026260932D-07 \\
  31    &0.77214443895274D-04  &0.13288355211454D-07 \\
  32    &0.14623108565057D-03  &0.14165730212821D-07 \\
\end{tabular}
\\[1.0cm]
\begin{tabular}{ccccc}
$t-range$  & $\chi^2$/d.o.f. & ${\bf{k}}^2$ &  $E({\bf{k}})$ &
$m_{\text{TS}}({\bf{k}})$  \\
\hline
0-32    &0.589     & 0.375  & 0.6382 (1) &  0.2142 (3)   \\
1-31    &0.377     & 0.375  & 0.6380 (2) &  0.2135 (5)    \\
2-30    &0.382     & 0.375  & 0.6381 (3) &  0.2139 (8)    \\
3-29    &0.328     & 0.375  & 0.6384 (4) &  0.2149 (13)     \\
\end{tabular}
\caption{The raw correlator data obtained for a spatial momentum
${\bf{k}}^2=0.375$ (corresponding to the integer assignment
$(n_x,n_y,n_z)=$ (0,3,1) on a $32^4$ lattice), for the value
$\kappa=0.074$ in the symmetric phase. The statistics is 135Ksweeps.
I also show the quality of the fits, the effective single-particle
energy and the time-slice mass obtained by using Eqs.(\ref{fitcor2})
and (\ref{disp3}) in various ranges of $t$.} \label{Table2}
\end{table}

\begin{table}
\begin{tabular}{ccc}
$t$  & $C_1(t,0;{\mathbf k})$ &  ${\text{statistical error}}$ \\
\hline
   0    &0.37264797340762D-03  &0.39868629339350D-07 \\
   1    &0.14423703753117D-03  &0.35128421137473D-07 \\
   2    &0.55852686582681D-04  &0.29204927503163D-07 \\
   3    &0.21643050470909D-04  &0.25324204561198D-07 \\
   4    &0.83911901659690D-05  &0.23695231347215D-07 \\
   5    &0.32553421665971D-05  &0.23752227353389D-07 \\
   6    &0.12625860175981D-05  &0.25644772176930D-07 \\
   7    &0.48959050096638D-06  &0.23531686001271D-07 \\
   8    &0.19263769002945D-06  &0.16045711212737D-07 \\
   9    &0.69459096205410D-07  &0.17742752632037D-07 \\
  10    &0.30668878114950D-07  &0.19019731595239D-07 \\
  11    &0.69459099077044D-07  &0.17742753057175D-07 \\
  12    &0.19263769455122D-06  &0.16045711766908D-07 \\
  13    &0.48959049739535D-06  &0.23531683326186D-07 \\
  14    &0.12625860091612D-05  &0.25644774746016D-07 \\
  15    &0.32553421642504D-05  &0.23752229326030D-07 \\
  16    &0.83911901631759D-05  &0.23695234653985D-07 \\
  17    &0.21643050344165D-04  &0.25324219112844D-07 \\
  18    &0.55852686485936D-04  &0.29204909466865D-07 \\
  19    &0.14423703585060D-03  &0.35128469121253D-07 \\
  20    &0.37264797340762D-03  &0.39868629339350D-07 \\
\end{tabular}
\\[1.0cm]
\begin{tabular}{ccccc}
$t-range$  & $\chi^2$/d.o.f. & ${\bf{k}}^2$ &  $E({\bf{k}})$ &
$m_{\text{TS}}({\bf{k}})$  \\
\hline
0-20    &0.451     & 0.922  & 0.9489 (1)  &  0.2186 (6)   \\
1-19    &0.274     & 0.922  & 0.9485 (3)  &  0.2162 (14)    \\
2-18    &0.244     & 0.922  & 0.9479 (6) &  0.2133 (33)    \\
3-17    &0.282     & 0.922  & 0.9476 (16) &  0.2120 (83)     \\
\end{tabular}
\caption{ The raw correlator data obtained for a spatial momentum
${\bf{k}}^2=0.922$ (corresponding to the integer assignment
$(n_x,n_y,n_z)=$ (3,1,0) on a $20^4$ lattice), for the value
$\kappa=0.074$ in the symmetric phase. The statistics is 215
Ksweeps. I also show the quality of the fits, the effective
single-particle energy and the time-slice mass obtained by using
Eqs.(\ref{fitcor2}) and (\ref{disp3}) in various ranges of $t$.}
\label{Table 3}
\end{table}

\begin{table}
\begin{tabular}{ccc}
$t$  & $C_1(t,0;{\mathbf k}=0)$ &  ${\text{statistical error}}$ \\
\hline
   0    &0.87514834087643D-03  &0.40185555074734D-05 \\
   1    &0.61452112729030D-03  &0.39930724185063D-05 \\
   2    &0.41664497529767D-03  &0.39573238547867D-05 \\
   3    &0.28249401086311D-03  &0.39576325347383D-05 \\
   4    &0.19223970120763D-03  &0.39509605466453D-05 \\
   5    &0.13173539366874D-03  &0.39095819880648D-05 \\
   6    &0.91507623807116D-04  &0.39253975622550D-05 \\
   7    &0.65276805609368D-04  &0.39656063145965D-05 \\
   8    &0.48943304332497D-04  &0.40437233918344D-05 \\
   9    &0.40073467468121D-04  &0.40949224665482D-05 \\
  10    &0.37300092795145D-04  &0.41228418720897D-05 \\
  11    &0.40073467468121D-04  &0.40949224665482D-05 \\
  12    &0.48943304332497D-04  &0.40437233918344D-05 \\
  13    &0.65276805609368D-04  &0.39656063145965D-05 \\
  14    &0.91507623807116D-04  &0.39253975622550D-05 \\
  15    &0.13173539366874D-03  &0.39095819880648D-05 \\
  16    &0.19223970120763D-03  &0.39509605466453D-05 \\
  17    &0.28249401086311D-03  &0.39576325347383D-05 \\
  18    &0.41664497529767D-03  &0.39573238547867D-05 \\
  19    &0.61452112729030D-03  &0.39930724185063D-05 \\
  20    &0.87514834087643D-03  &0.40185555074734D-05 \\
\end{tabular}
\\[1.0cm]
\begin{tabular}{ccccc}
$t-range$  & $\chi^2$/d.o.f. & ${\bf{k}}^2$ &  $E({\bf{k}})$ &
$m_{\text{TS}}({\bf{k}})$  \\
\hline
0-20    &1.572     & 0.000  & 0.3804 (17) &  0.3827 (17)   \\
1-19    &0.005     & 0.000  & 0.3895 (24) &  0.3920 (24)    \\
2-18    &0.005     & 0.000  & 0.3892 (35) &  0.3917 (35)    \\
3-17    &0.002     & 0.000  & 0.3884 (50) &  0.3909 (51)     \\

\end{tabular}
\caption{ The raw correlator data, at ${\bf{k}}=0$ and for
$\kappa=0.076$ in the broken phase. The statistics is 6 Msweeps on a
$20^4$ lattice. I also show the quality of the fits with
Eq.(\ref{fitcor2}), the effective single particle energy and the
time-slice mass obtained in various ranges of $t$. The vacuum
expectation value was $\langle|\phi|\rangle= 0.30157(3)$ and the
susceptibility $\chi_{\text{latt}}=37.87(9)$~.} \label{Table 4}
\end{table}

\begin{table}
\begin{tabular}{ccc}
$t$  & $C_1(t,0;{\mathbf k})$ &  ${\text{statistical error}}$ \\
\hline
   0    &0.71526729630726D-03  &0.13615402429507D-05 \\
   1    &0.43018990381477D-03  &0.13115919558418D-05 \\
   2    &0.26023053854951D-03  &0.12910546778340D-05 \\
   3    &0.15766111080779D-03  &0.12742350485637D-05 \\
   4    &0.95697663109043D-04  &0.12054623450829D-05 \\
   5    &0.58281796633920D-04  &0.11691774644051D-05 \\
   6    &0.35714557269430D-04  &0.10649585762136D-05 \\
   7    &0.22277393587883D-04  &0.99046217772169D-06 \\
   8    &0.14385158394354D-04  &0.10404673826236D-05 \\
   9    &0.10544242495683D-04  &0.11622441471229D-05 \\
  10    &0.94058521329496D-05  &0.13030286785723D-05 \\
  11    &0.10544242398532D-04  &0.11622441580206D-05 \\
  12    &0.14385158341362D-04  &0.10404674051829D-05 \\
  13    &0.22277393469260D-04  &0.99046212649583D-06 \\
  14    &0.35714557356030D-04  &0.10649585835496D-05 \\
  15    &0.58281796639610D-04  &0.11691775666978D-05 \\
  16    &0.95697663247686D-04  &0.12054621641341D-05 \\
  17    &0.15766111116237D-03  &0.12742352704367D-05 \\
  18    &0.26023053826732D-03  &0.12910544561123D-05 \\
  19    &0.43018990454806D-03  &0.13115923352524D-05 \\
  20    &0.71526729630726D-03  &0.13615402429507D-05 \\
\end{tabular}
\\[1.0cm]
\begin{tabular}{ccccc}
$t-range$  & $\chi^2$/d.o.f. & ${\bf{k}}^2$ &  $E({\bf{k}})$ &
$m_{\text{TS}}({\bf{k}})$  \\
\hline
0-20    &0.235     & 0.098  & 0.5044 (10) & 0.4024 (13)   \\
1-19    &0.011     & 0.098  & 0.5019 (16) & 0.3992 (21)    \\
2-18    &0.007     & 0.098  & 0.5014 (25) & 0.3985 (32)    \\
3-17    &0.009     & 0.098  & 0.5014 (38) & 0.3985 (49)     \\
\end{tabular}
\caption{The raw correlator data for $\kappa=0.076$ in the broken
phase, on a $20^4$ lattice, for the integer assignment (1,0,0). The
statistics is 230 Ksweeps. I also show the quality of the fits, the
effective single-particle energy and the time-slice mass obtained in
various ranges of $t$.} \label{Table 5}
\end{table}

\begin{table}
\begin{tabular}{ccc}
$t$  & $C_1(t,0;{\mathbf k})$ &  ${\text{statistical error}}$ \\
\hline
   0    &0.60134431492201D-03  &0.72278071460665D-06 \\
   1    &0.33121256195148D-03  &0.67224614280968D-06 \\
   2    &0.18340753552086D-03  &0.57001952486459D-06 \\
   3    &0.10169038556086D-03  &0.58126295776147D-06 \\
   4    &0.56548852939487D-04  &0.62601000175195D-06 \\
   5    &0.31288412716245D-04  &0.61885213006675D-06 \\
   6    &0.17273284231833D-04  &0.61847800661692D-06 \\
   7    &0.98445753758052D-05  &0.60619584062421D-06 \\
   8    &0.58757929618968D-05  &0.64071878068011D-06 \\
   9    &0.40294379323518D-05  &0.68994571329391D-06 \\
  10    &0.34202276827878D-05  &0.74746648254179D-06 \\
  11    &0.40294380412313D-05  &0.68994574565471D-06 \\
  12    &0.58757929936507D-05  &0.64071876223713D-06 \\
  13    &0.98445753579537D-05  &0.60619580358585D-06 \\
  14    &0.17273284284194D-04  &0.61847800602069D-06 \\
  15    &0.31288412838347D-04  &0.61885209771265D-06 \\
  16    &0.56548852916696D-04  &0.62600984641147D-06 \\
  17    &0.10169038565092D-03  &0.58126288128429D-06 \\
  18    &0.18340753431478D-03  &0.57001966754573D-06 \\
  19    &0.33121256136276D-03  &0.67224582018553D-06 \\
  20    &0.60134431492201D-03  &0.72278071460665D-06 \\
\end{tabular}
\\[1.0cm]
\begin{tabular}{ccccc}
$t-range$  & $\chi^2$/d.o.f. & ${\bf{k}}^2$ &  $E({\bf{k}})$ &
$m_{\text{TS}}({\bf{k}})$  \\
\hline
0-20    &0.358     & 0.196  & 0.5932 (8) &  0.4081 (12)   \\
1-19    &0.036     & 0.196  & 0.5905 (14) &  0.4039 (21)    \\
2-18    &0.036     & 0.196  & 0.5899 (23) &  0.4030 (35)    \\
3-17    &0.042     & 0.196  & 0.5902 (42) &  0.4034 (63)     \\
\end{tabular}
\caption{ The raw correlator data for $\kappa=0.076$ in the broken
phase, on a $20^4$ lattice, for the integer assignment (1,1,0). The
statistics is 230 Ksweeps. I also show the quality of the fits, the
effective single-particle energy and the time-slice mass obtained in
various ranges of $t$.} \label{Table 6}
\end{table}

\begin{table}
\begin{tabular}{ccc}
$t$  & $C_1(t,0;{\mathbf k})$ &  ${\text{statistical error}}$ \\
\hline
   0  &  0.34053202765797D-03 & 0.25796702821673D-07\\
   1  &  0.12939893707440D-03 & 0.22644567310486D-07\\
   2  & 0.49445741636059D-04  & 0.18855019661045D-07\\
   3  &  0.18941908006539D-04 & 0.18965679348131D-07\\
   4  &  0.72475836118000D-05 & 0.18733026319633D-07\\
   5  &  0.27696225381887D-05 & 0.19177920168895D-07\\
   6  &  0.10512924615448D-05 & 0.18318892429225D-07\\
   7  &  0.41757228168673D-06 & 0.18213926845448D-07\\
   8  &  0.17889655811547D-06 & 0.21653248311599D-07\\
   9  &  0.75160519493844D-07 & 0.26073999719728D-07\\
  10  &  0.37277101746356D-07 & 0.31392903787941D-07\\
  11  &  0.75160519493844D-07 & 0.26073999719728D-07\\
  12  &  0.17889655811547D-06 & 0.21653248311599D-07\\
  13  &  0.41757228168673D-06 & 0.18213926845448D-07\\
  14  &  0.10512924615448D-05 & 0.18318892429225D-07\\
  15  &  0.27696225381887D-05 & 0.19177920168895D-07\\
  16  &  0.72475836118000D-05 & 0.18733026319633D-07\\
  17  &  0.18941908006539D-04 & 0.18965679348131D-07\\
  18  &  0.49445741636059D-04 & 0.18855019661045D-07\\
  19  &  0.12939893707440D-03 & 0.22644567310486D-07\\
  20  &  0.34053202765797D-03 & 0.25796702821673D-07\\
\end{tabular}
\\[1.0cm]
\begin{tabular}{ccccc}
$t-range$  & $\chi^2$/d.o.f. & ${\bf{k}}^2$ &  $E({\bf{k}})$ &
$m_{\text{TS}}({\bf{k}})$  \\
\hline
0-20    &25.51   & 0.824  & 0.9656(1)   &  0.4275 (2)   \\
1-19    &0.794    & 0.824  & 0.9614 (2)   &  0.4163 (6)    \\
2-18    &0.278     & 0.824  & 0.9599 (6)  &  0.4122 (15)    \\
3-17    &0.272     & 0.824  & 0.9609 (15)  &  0.4150 (40)     \\
\end{tabular}
\caption{ The raw correlator data for $\kappa=0.076$ in the broken
phase, on a $20^4$ lattice, for the integer assignment (3,0,0). The
statistics is 10 Msweeps. I also show the quality of the fits, the
effective single-particle energy and the time-slice mass obtained in
various ranges of $t$.} \label{Table 7}
\end{table}

\begin{table}
\begin{tabular}{ccc}
$t$  & $C_1(t,0;{\mathbf k})$ &  ${\text{statistical error}}$ \\
\hline
   0 &   0.32191840093250D-03 & 0.28215740200667D-07\\
   1 &   0.11716187697558D-03 & 0.24420632370780D-07\\
   2 &   0.42874548504809D-04 & 0.19461125127975D-07\\
   3 &   0.15726715346676D-04 & 0.17170759460693D-07\\
   4 &   0.57725432712762D-05 & 0.17245120570366D-07\\
   5 &   0.21143021635064D-05 & 0.17637391271899D-07\\
   6 &   0.77592919860017D-06 & 0.17545895751891D-07\\
   7 &   0.29550904531463D-06 & 0.17673166191011D-07\\
   8 &   0.12817308964179D-06 & 0.18301057235211D-07\\
   9 &   0.57731193876899D-07 & 0.23297882913256D-07\\
  10 &   0.29487922332530D-07 & 0.25758084311071D-07\\
  11 &   0.57731193876899D-07 & 0.23297882913256D-07\\
  12 &   0.12817308964179D-06 & 0.18301057235211D-07\\
  13 &   0.29550904531463D-06 & 0.17673166191011D-07\\
  14 &   0.77592919860017D-06 & 0.17545895751891D-07\\
  15 &   0.21143021635064D-05 & 0.17637391271899D-07\\
  16 &   0.57725432712762D-05 & 0.17245120570366D-07\\
  17 &   0.15726715346676D-04 & 0.17170759460693D-07\\
  18 &   0.42874548504809D-04 & 0.19461125127975D-07\\
  19 &   0.11716187697558D-03 & 0.24420632370780D-07\\
  20 &   0.32191840093250D-03 & 0.28215740200667D-07\\
\end{tabular}
\\[1.0cm]
\begin{tabular}{ccccc}
$t-range$  & $\chi^2$/d.o.f. & ${\bf{k}}^2$ &  $E({\bf{k}})$ &
$m_{\text{TS}}({\bf{k}})$  \\
\hline
0-20    &19.00    & 0.922  & 1.0087 (1)  &  0.4295 (3)   \\
1-19    &0.858    & 0.922  & 1.0046 (3)  &  0.4178 (7)    \\
2-18    &0.309     & 0.922  & 1.0028 (6) &  0.4127 (18)    \\
3-17    &0.351     & 0.922  & 1.0022 (17) &  0.4113 (48)     \\
\end{tabular}
\caption{ The raw correlator data for $\kappa=0.076$ in the broken
phase, on a $20^4$ lattice, for the integer assignment (3,1,0). The
statistics is  10 Msweeps. I also show the quality of the fits, the
effective single-particle energy and the time-slice mass obtained in
various ranges of $t$.} \label{Table 8}
\end{table}

\begin{table}
\begin{tabular}{ccc}
$t$  & $C_1(t,0;{\mathbf k})$ &  ${\text{statistical error}}$ \\
\hline
   0 &   0.30566681793518D-03 & 0.24505540691796D-07\\
   1 &   0.10682582826270D-03 & 0.20658932104437D-07\\
   2 &   0.37545987045166D-04 & 0.15460405900989D-07\\
   3 &   0.13225370098169D-04 & 0.14067447764431D-07\\
   4 &   0.46577041941379D-05 & 0.15233371727808D-07\\
   5 &   0.16509226154241D-05 & 0.16965416047581D-07\\
   6 &   0.58929799490177D-06 & 0.17141230984827D-07\\
   7 &   0.20719084545598D-06 & 0.15254671198613D-07\\
   8 &   0.69295505341081D-07 & 0.15879409375053D-07\\
   9 &   0.22200695571408D-07 & 0.17865514829934D-07\\
  10 &   0.16786349040784D-07 & 0.20544843933658D-07\\
  11 &   0.22200695571408D-07 & 0.17865514829934D-07\\
  12 &   0.69295505341081D-07 & 0.15879409375053D-07\\
  13 &   0.20719084545598D-06 & 0.15254671198613D-07\\
  14 &   0.58929799490177D-06 & 0.17141230984827D-07\\
  15 &   0.16509226154241D-05 & 0.16965416047581D-07\\
  16 &   0.46577041941379D-05 & 0.15233371727808D-07\\
  17 &   0.13225370098169D-04 & 0.14067447764431D-07\\
  18 &   0.37545987045166D-04 & 0.15460405900989D-07\\
  19 &   0.10682582826270D-03 & 0.20658932104437D-07\\
  20 &   0.30566681793518D-03 & 0.24505540691796D-07\\
\end{tabular}
\\[1.0cm]
\begin{tabular}{ccccc}
$t-range$  & $\chi^2$/d.o.f. & ${\bf{k}}^2$ &  $E({\bf{k}})$ &
$m_{\text{TS}}({\bf{k}})$  \\
\hline
0-20    &21.23    & 1.020  & 1.0493 (1)  &  0.4308 (3)   \\
1-19    &0.667     & 1.020  & 1.0450 (2)  &  0.4183 (7)    \\
2-18    &0.135     & 1.020  & 1.0432 (6) &  0.4129 (20)    \\
3-17    &0.131     & 1.020  & 1.0422 (20) &  0.4099 (61)     \\
\end{tabular}
\caption{ The raw correlator data for $\kappa=0.076$ in the broken
phase, on a $20^4$ lattice, for the integer assignment (3,1,1). The
statistics is 10 Msweeps. I also show the quality of the fits, the
effective single-particle energy and the time-slice mass obtained in
various ranges of $t$.} \label{Table 9}
\end{table}

\begin{table}
\begin{tabular}{ccc}
$t$  & $C_1(t,0;{\mathbf k}=0)$ &  ${\text{statistical error}}$ \\
\hline
   0  &  0.18825024647556D-03 & 0.16941891466779D-05\\
   1  & 0.14042050173424D-03  & 0.16936167168555D-05\\
   2  &  0.98762414233733D-04 & 0.16927933487534D-05\\
   3  &  0.68288037480761D-04 & 0.16919086215327D-05\\
   4  &  0.46981340861454D-04 & 0.16913027902220D-05\\
   5  &  0.32320623399664D-04 & 0.16907071684861D-05\\
   6  &  0.22322431762012D-04 & 0.16901776642593D-05\\
   7  &  0.15515610279490D-04 & 0.16897524471058D-05\\
   8  &  0.10880836739844D-04 & 0.16894000480709D-05\\
   9  &  0.77275164550180D-05 & 0.16893186514232D-05\\
  10  &  0.55921686139766D-05 & 0.16891971059864D-05\\
  11  &  0.41522126343909D-05 & 0.16891265663505D-05\\
  12  &  0.31962571843670D-05 & 0.16893317502024D-05\\
  13  &  0.25821007487241D-05 & 0.16897826971614D-05\\
  19  &  0.25821007487241D-05 & 0.16897826971614D-05\\
  20  &  0.31962571843670D-05 & 0.16893317502024D-05\\
  21  &  0.41522126343909D-05 & 0.16891265663505D-05\\
  22  &  0.55921686139766D-05 & 0.16891971059864D-05\\
  23  &  0.77275164550180D-05 & 0.16893186514232D-05\\
  24  &  0.10880836739844D-04 & 0.16894000480709D-05\\
  25  &  0.15515610279490D-04 & 0.16897524471058D-05\\
  26  &  0.22322431762012D-04 & 0.16901776642593D-05\\
  27  &  0.32320623399664D-04 & 0.16907071684861D-05\\
  28  &  0.46981340861454D-04 & 0.16913027902220D-05\\
  29  &  0.68288037480761D-04 & 0.16919086215327D-05\\
  30  &  0.98762414233733D-04 & 0.16927933487534D-05\\
  31  &  0.14042050173424D-03 & 0.16936167168555D-05\\
  32  &  0.18825024647556D-03 & 0.16941891466779D-05\\
\end{tabular}
\caption{The raw correlator data at ${\bf{k}}=0$ obtained from a
simulation on a $32^4$ lattice, for $\kappa=0.076$ in the broken
phase. The statistics is 3500 Ksweeps. Only data with $S/N >1$ are
reported. The vacuum expectation value was $\langle|\phi|\rangle=0.30158(2)$ and the susceptibility $\chi_{\text{latt}}= 37.70(11)$.}
\label{Table 10}
\end{table}

\begin{table}
\begin{tabular}{ccc}
$t$  & $C_1(t,0;{\mathbf k}=0)$ &  ${\text{statistical error}}$ \\
\hline
   0 &   0.19159658349118D-03 & 0.42089126114641D-05 \\
   1  &  0.14376486787758D-03 & 0.42064742520006D-05  \\
   2  &  0.10210247213207D-03 &  0.41965516984451D-05  \\
   3  &  0.71577996915195D-04 & 0.41737044386843D-05  \\
   4  &  0.50219314163541D-04 & 0.41429253869347D-05  \\
   5  &  0.35550394123410D-04 & 0.40976377585245D-05  \\
   6  &  0.25544524377202D-04 & 0.40649605084329D-05  \\
   7  &  0.18747584437939D-04 & 0.40363212212057D-05  \\
   8  &  0.14150039767838D-04 & 0.40174857064957D-05  \\
   9  &  0.11028271382665D-04 & 0.40038139588244D-05  \\
  10  &  0.88824632819617D-05 & 0.39929623745097D-05  \\
  11  &  0.74336822982801D-05 & 0.39881786206720D-05  \\
  12  &  0.64935569222907D-05 & 0.40024501405502D-05  \\
  13  &  0.58917012270399D-05 & 0.40348389009236D-05  \\
  19  &  0.58917012270399D-05 & 0.40348389009236D-05  \\
  20  &  0.64935569222907D-05 & 0.40024501405502D-05  \\
  21  &  0.74336822982801D-05 & 0.39881786206720D-05  \\
  22  &  0.88824632819617D-05 & 0.39929623745097D-05  \\
  23  &  0.11028271382665D-04 & 0.40038139588244D-05  \\
  24  &  0.14150039767838D-04 & 0.40174857064957D-05  \\
  25  &  0.18747584437939D-04 & 0.40363212212057D-05  \\
  26  &  0.25544524377202D-04 & 0.40649605084329D-05  \\
  27  &  0.35550394123410D-04 & 0.40976377585245D-05 \\
  28  &  0.50219314163541D-04 & 0.41429253869347D-05 \\
  29  &  0.71577996915195D-04 & 0.41737044386843D-05  \\
  30  &  0.10210247213207D-03 & 0.41965516984451D-05  \\
  31  &  0.14376486787758D-03 & 0.42064742520006D-05  \\
  32  &  0.19159658349118D-03 & 0.42089126114641D-05 \\
\end{tabular}
\caption{The raw correlator data at ${\bf{k}}=0$ obtained from a
simulation on a $32^4$ lattice, for $\kappa=0.076$ in the broken
phase. The statistics is 1250Ksweeps. Only data with $S/N >1$ are
reported. The vacuum expectation value was $\langle|\phi|\rangle=0.30156(4)$ and the susceptibility $\chi_{\text{latt}}= 37.72(18)$.}
\label{Table 11}
\end{table}

\begin{table}
\begin{tabular}{ccc}
$t$  & $C_1(t,0;{\mathbf k}=0)$ &  ${\text{statistical error}}$ \\
\hline
   0 &   0.18830816108376D-03 & 0.36295412301977D-05 \\
   1 &   0.14049039900582D-03 & 0.36347914715019D-05\\
   2 &   0.98868629477251D-04 & 0.36448907002102D-05\\
   3 &   0.68418079591984D-04 & 0.36545903093631D-05\\
   4 &   0.47126928040045D-04 & 0.36613388157817D-05\\
   5 &   0.32499368020916D-04 & 0.36675738973308D-05\\
   6 &   0.22525563622636D-04 & 0.36739990211847D-05\\
   7 &   0.15786687380170D-04 & 0.36845952758579D-05\\
   8 &   0.11231440928165D-04 & 0.37072940633664D-05\\
   9 &   0.81636421519316D-05 & 0.37271696019070D-05\\
  10 &   0.60622792191850D-05 & 0.37404570700649D-05\\
  11 &   0.46039724904828D-05 & 0.37442837585148D-05\\
  21 &   0.46039724904828D-05 & 0.37442837585148D-05\\
  22 &   0.60622792191850D-05 & 0.37404570700649D-05\\
  23 &   0.81636421519316D-05 & 0.37271696019070D-05\\
  24 &   0.11231440928165D-04 & 0.37072940633664D-05\\
  25 &   0.15786687380170D-04 & 0.36845952758579D-05\\
  26 &   0.22525563622636D-04 & 0.36739990211847D-05\\
  27 &   0.32499368020916D-04 & 0.36675738973308D-05\\
  28 &   0.47126928040045D-04 & 0.36613388157817D-05\\
  29 &   0.68418079591984D-04 & 0.36545903093631D-05\\
  30 &   0.98868629477251D-04 & 0.36448907002102D-05\\
  31 &   0.14049039900582D-03 & 0.36347914715019D-05\\
  32 &   0.18830816108376D-03 & 0.36295412301977D-05\\
\end{tabular}
\caption{The raw correlator data at ${\bf{k}}=0$ obtained from a
simulation on a $32^4$ lattice, for $\kappa=0.076$ in the broken
phase. The statistics is 820Ksweeps. Only data with $S/N >1$ are
reported. The vacuum expectation value was $\langle|\phi|\rangle=0.30157(3)$ and the susceptibility $\chi_{\text{latt}}= 37.74(20)$.}
\label{Table 12}
\end{table}

\begin{table}
\begin{tabular}{ccc}
$t$  & $C_1(t,0;{\mathbf k}=0)$ &  ${\text{statistical error}}$ \\
\hline

   0 &   0.18662641483139D-03 & 0.42499108352562D-05 \\
   1 &   0.13880114074366D-03 & 0.42356693269278D-05\\
   2 &   0.97166753677375D-04 & 0.42277156923244D-05\\
   3 &   0.66681534205013D-04 & 0.42123078274730D-05\\
   4 &   0.45350298378516D-04 & 0.42041206560544D-05\\
   5 &   0.30654197711880D-04 & 0.42124873432185D-05\\
   6 &   0.20607943469257D-04 & 0.42459030224462D-05\\
   7 &   0.13757407397773D-04 & 0.42864246337440D-05\\
   8 &   0.90869249084038D-05 & 0.43060940267945D-05\\
   9 &   0.59327998437900D-05 & 0.43120683752934D-05\\
  23 &   0.59327998437900D-05 & 0.43120683752934D-05\\
  24 &   0.90869249084038D-05 & 0.43060940267945D-05\\
  25 &   0.13757407397773D-04 & 0.42864246337440D-05\\
  26 &   0.20607943469257D-04 & 0.42459030224462D-05\\
  27 &   0.30654197711880D-04 & 0.42124873432185D-05\\
  28 &   0.45350298378516D-04 & 0.42041206560544D-05\\
  29 &   0.66681534205013D-04 & 0.42123078274730D-05\\
  30 &   0.97166753677375D-04 & 0.42277156923244D-05\\
  31 &   0.13880114074366D-03 & 0.42356693269278D-05\\
  32 &   0.18662641483139D-03 & 0.42499108352562D-05\\
\end{tabular}
\caption{The raw correlator data at ${\bf{k}}=0$ from a simulation
on a $32^4$ lattice, for $\kappa=0.076$ in the broken phase. The
statistics is 620Ksweeps. Only data with $S/N>1$ are reported. The
vacuum expectation value was $\langle|\phi|\rangle= 0.30157(3)$ and
the susceptibility $\chi_{\text{latt}}= 37.72(28)$.} \label{Table
13}
\end{table}

\begin{table}
\begin{tabular}{ccccc}
$t-range$  & $\chi^2$/d.o.f. & ${\bf{k}}^2$ &  $E({\bf{k}})$ &
$m_{\text{TS}}({\bf{k}})$  \\
\hline
0-32    &0.779     & 0.000  & 0.3443 (25) &  0.3460 (25)   \\
1-31    &0.255     & 0.000  & 0.3624 (37) &  0.3644 (37)    \\
2-30    &0.254     & 0.000  & 0.3655 (54) &  0.3675 (54)    \\
3-29    &0.256     & 0.000  & 0.3622 (77) &  0.3642 (77)     \\
4-28    &0.252     & 0.000  & 0.3546 (110)&  0.3565 (110)    \\
\end{tabular}
\caption{ The results of the global fit to the correlator data
reported in Tables 10-13, for $\kappa=0.076$ in the broken phase.
The total statistics is 6.19 Msweeps on  $32^4$ lattices. I report
the quality of the fits with Eq.(\ref{fitcor2}), the effective
single particle energy and the time-slice mass obtained in various
ranges of $t$. These results should be compared with those in Table
4 from the $20^4$ lattice. } \label{Table 14}
\end{table}

\begin{table}
\begin{tabular}{ccccc}
$t-range$  & $\chi^2$/d.o.f. & ${\bf{k}}^2$ &  $E_1({\bf{k}})$ &
$m_{\text{TS}}({\bf{k}})$  \\
\hline
0-20    &1.573    & 0.000  & 0.3804 (17) &  0.3827 (17)   \\
1-19    &0.002     & 0.000  & 0.3880 (71) &  0.3904 (71)    \\
2-18    &0.001     & 0.000  & 0.3866 (113) &  0.3890 (113)    \\
3-17    &0.000    & 0.000  & 0.3861 (159) &  0.3885 (159)     \\
\end{tabular}
\caption{ The results of a fit to the correlator data reported in
Table 4 from the $20^4$ lattice. The fit is performed using a
constrained 2-mass formula with $E_2(0)=2E_1(0)$ as suggested in
Ref.\cite{balog2}. I report the quality of the fits, $E_1(0)$ and
the time-slice mass obtained in various ranges of $t$. These results
should be compared with those in Table 4 obtained fitting the
correlator data to Eq.(\ref{fitcor2}). } \label{Table 15}
\end{table}

\begin{table}
\begin{tabular}{ccccc}
$t-range$  & $\chi^2$/d.o.f. & ${\bf{k}}^2$ &  $E_1({\bf{k}})$ &
$m_{\text{TS}}({\bf{k}})$  \\
\hline
0-32    &0.787     & 0.000  & 0.3443 (25) &  0.3460 (25)   \\
1-31    &0.258     & 0.000  & 0.3624 (37) &  0.3644 (37)    \\
2-30    &0.236     & 0.000  & 0.3416 (198) &  0.3433 (198)    \\
3-29    &0.225    & 0.000  & 0.3143 (324) &  0.3156 (324)     \\

\end{tabular}
\caption{The results of a fit to the correlator data reported in
Tables 10-13 from the $32^4$ lattice. The fit is performed using a
constrained 2-mass formula with $E_2(0)=2E_1(0)$ as suggested in
Ref.\cite{balog2}. I report the quality of the fits, $E_1(0)$ and
the time-slice mass obtained in various ranges of $t$. These results
should be compared with those in Table 14 obtained fitting the same
correlator data with Eq.(\ref{fitcor2}) and with those in Table 15
obtained from the $20^4$ lattice data using the same 2-mass fitting
function.} \label{Table 16}
\end{table}

\end{document}